\documentstyle[aps,12pt]{revtex}

\begin{document}
\title{Electronic structure of the V$^{3+}$ ion in V$_{2}$O$_{3}^{\ast }$ }
\author{R.J. Radwanski}
\address{Center for Solid State Physics, S$^{nt}$ Filip 5,31-150Krakow,Poland%
\\
Institute of Physics, Pedagogical University, 30-084 Krakow, Poland}
\author{Z. Ropka}
\address{Center for Solid State Physics, S$^{nt}$ Filip\\
5,31-150Krakow,Poland.\\
e-mail: sfradwan@cyf-kr.edu.pl; http://www.css-physics.edu.pl}
\maketitle

\begin{abstract}
We have attributed magnetism and electronic structure of V$_{2}$O$_{3}$ to
the V$^{3+}$ ions. We claim that the V$^{3+}$ ion in V$_{2}$O$_{3}$ should
be considered as described by the quantum numbers {\it S}=1 and {\it L}=3.
Such quantum numbers result from two Hund's rules. The resulting electronic
structure is much more complex than considered up to now, but it describes,
e.g. the insulating ground state and the lowered magnetic moment, of 1.2 $%
\mu _{B}$. It turns out that the intra-atomic spin-orbit coupling is
indispensable for the physically adequate description of electronic and
magnetic properties of V$_{2}$O$_{3}$.

Keywords: highly-correlated electron system, crystal field, V$^{3+}$ ion,
spin-orbit coupling, Mott insulators

PACS: 71.70.E, 75.10.D, 75.30.Gw
\end{abstract}

\pacs{71.70.E, 75.10.D, 75.30.Gw;}
\date{(11.03.2003)}

\section{Introduction}

V$_{2}$O$_{3}$ attracts the substantial scientific interest by more than 50
years \ \cite{1}. Despite of it there is still strong discussion about the
description of its properties and its electronic structure. There is a
long-standing controversy between a {\it S}=1 model without an orbital
degeneracy \cite{2} and the $S$=1/2 orbitally degenerate model of Castellani
et al.\cite{3}. Recently an ''orbitally degenerate spin-1 model for the
insulating V$_{2}$O$_{3}$'' has been proposed in Ref. \cite{4,5} whereas
spin-1 model with three degenerate orbitals quite recently was worked out by
Di Matteo \cite{6}.

In all of these considerations it is agreed that V$_{2}$O$_{3}$ is an
insulating antiferromagnet with the Neel temperature of 160 K. At that
temperature also the low-temperature insulating state transforms to a
metallic state without long-range magnetic order. V$_{2}$O$_{3}$ has
basically the rhombohedral corundum Al$_{2}$O$_{3}$ structure. It is agreed
that V ions in the corundum structure sit predominantly in the oxygen
octahedron with some further distortions, Fig. 1 of Ref. \cite{4,5}. It is
also agreed that the V$^{3+}$ ion has two {\it d} electrons, i.e. that the
charge fluctuations are practically absent \cite{2,3,4}. The basis for all
theories is the description of the V$^{3+}$ ion and its electronic structure.

The aim of this paper is to point out that the V$^{3+}$ ion in V$_{2}$O$_{3}$
should be described by the quantum numbers $S$=1 and $L$=3.

\section{Theoretical outline}

The V$^{3+}$ ion has two 3$d$ electrons. According to us these two electrons
form the highly-correlated atomic-like electronic system 3$d^{n}$ ($n$=2)
that is described by the resultant quantum numbers $S$=1 and $L$=3. Its
electronic structure contains 21 discrete states. According to the Quantum
Atomistic Solid State Theory, QUASST, these states are preserved also when
the paramagnetic ion becomes the full part of a solid \cite{7}. The
resultant quantum numbers $S$=1, $L$=3 and $^{3}F$ ground term, Fig. 1a,
come out from two Hund's rules (two, i.e. 1$^{o}$ the maximal $S$ and 2$^{o}$
the maximal $L$ for two $d$ electrons). Hund's rules reflect strong
intra-atomic interactions. Thus, the demand for keeping the Hund rules is
equivalent to an idea that the paramagnetic atom preserves much of their
internal atomic structure being the part of a solid. Such the atomic-like 3$%
d^{n}$ system interacts with the charge and spin surrounding in the solid.
The interaction with the charge surrounding we approximate by means of the
crystal-field interactions. The symmetry of the crystal-field interactions
depends on the local symmetry and in the corrundum structure is
predominantly octahedral. The calculated discrete electronic structure in
the octahedral crystal field, Fig. 1b, and in the presence of the
intra-atomic spin-orbit coupling is shown in Fig. 1c. To such electronic
structure we superimpose the spin-dependent interactions to account for the
magnetic state. The self-consistent calculations are performed similarly to
that presented in Ref. \cite{8} for FeBr$_{2}$. Having the electronic
structure, both in magnetic and paramagnetic state, we can calculate the
free Helmholtz energy and the resulting thermodynamical properties by means
of the statistical physics.

\section{Results and discussion}

The shown energy level scheme of the 3$d^{2}$ system in the octahedral
crystal field and in the presence of the spin-orbit coupling, with the
octahedral crystal field dominating the intra-atomic spin-orbit coupling,
contains 21 states. As seen from Fig. 1c the dominating octahedral crystal
field leaves 9 lowest states well separated from others. These 9 states are
spread over 60 meV and their eigenfunctions and energies determine the
detailed properties at room and lower temperatures. The properties at zero
temperature are determined by the properties of the ground state. The
detailed splitting depends on value of the spin-orbit coupling and on the
distortions.

The ground state is an accidental non-magnetic doublet. This doublet
structure is easily split by means of lattice distortions and/or
spin-dependent interactions. They yield a singlet ground state of the V$%
^{3+} $ ion in the atomic scale. Despite of the single-ion singlet ground
state the magnetism can easily develop as the closely lying states work as a
pseudo-doublet. The similar moment-induced mechanism is well known in
praseodymium compounds \cite{9}. As the result of local distortions the
local V moment can be modelled, both its value and the direction. We get a
value of 1.2 $\mu _{B}$ for the V$^{3+}$-ion magnetic moment in the ordered
state at 0 K - this value well reproduces the experimental value. We use the
octahedral CEF parameter $B_{4}^{0}$ = -40 K and the spin-orbit coupling $%
\lambda _{s-o}$=+150 K. The sign of $B_{4}$ comes out directly from the
calculations of the octupolar potential for the octahedral oxygen-anion
surroundings.

Our approach yields in the very natural way the insulating ground state. Due
to strong correlations all electrons after the charge transfer, from the
metal atoms to oxygens, during the formation of the compound, are localized
at the oxygens. Within our approach we can understand why the cubic
perovskite structure distorts. By the distortion the system gains energy.
From 1937, after Jahn and Teller's work we know that system will
spontaneously distort to remove the accidental double degeneracy of the
ground state. Our understanding of 3$d$-ion compounds is close to an
original idea of Van Vleck from 1932, that electronic and magnetic
properties are largely determined by the atomic-like electronic structure.

\section{Conclusions}

We have attributed magnetism and electronic structure of V$_{2}$O$_{3}$ to
the V$^{3+}$ ions. We claim that the V$^{3+}$ ion in V$_{2}$O$_{3}$ should
be considered as described by the quantum numbers $S$=1 and $L$=3. Such
quantum numbers result from two Hund's rules. We are convinced that the
orbital degree of freedom, related to the orbital quantum number $L$ and the
intra-atomic spin-orbit coupling are indispensable for the physically
adequate description of electronic and magnetic properties of V$_{2}$O$_{3}$%
. Most up-to-now approaches employ the one-electron $t_{2g}$ (ground state)
and excited $e_{g}$ orbitals and next consider a molecular V-V unit. Within
the Quantum Atomistic Solid-State theory we have shown that the value and
the direction of the local magnetic moment is predominantly governed by the
local lattice symmetry, e.g. Ref. \cite{6}. We get a value of 1.2 $\mu _{B}$
for the V$^{3+}$ ion magnetic moment - this value is very sensitive to the
local distortions and contains the substantial orbital moment. We are
convinced that our description is also useful for the description of the V$%
^{3+}$ ions in orthovanades LaVO$_{3}$ and YVO$_{3}$ \cite{10}. This paper
has been at first rejected due to ''an unusual theoretical concept'' in the
situation where virtually all other workers in the field over the last
twenty years have adopted the V-V molecular unit in the unit cell as a basis
for their discussion.'' For us, the atomic approach as the start for the
analysis of a compound is obvious owing to a presently well-established, but
originally very doubtful, an idea of Dalton and Davy that all solids are
built from atoms. So different approaches assure the useful discussion and
we believe that the publishing of our paper enables the open scientific
discussion and allows to solve the problem in the future. As far as
correlations are regarding our approach to V$_{2}$O$_{3}$ is in the very
strongly-correlated limit, though these correlations are primarily on-site
and atomic-like. Developing of very different theories, despite of 20 years
of studies and in the very last time \cite{10,11,12}, we take as indication
that there is very urging need for novel theoretical approaches even such
''unusual'' as the one presented in this paper. However, we can add that our
model provides both the orbital and spin moment \cite{8,13}. We came out
with our approach to 3$d$-atom containing compounds in 1996 \cite{14} when
the orbital moment was not yet revealed experimentally. Thus, we take the
possibility of accounting for the experimentally-revealed orbital moment as
the great plus for our theoretical approach. But surely there is still a lot
to do in order to describe details of experimental details, but we feel that
the atomic-scale properties are essentially important for description of
magnetic and electronic properties of 3$d$-atom compounds.

* This paper has been submitted 31.05.2002 to Strongly Correlated Electron
Conference in Krakow, SCES-02 getting a code MOT026. It has been presented
at the Conference but has been rejected with referee's arguments that ''The
authors should make a convincing case for their adoption of a strictly
atomic model in the discussion of the physical properties of the V$^{3+}$
ion in V$_{2}$O$_{3}$ when virtually all other workers in the field over the
last twenty years have adopted the V-V molecular unit in the unit cell as a
basis for their discussion. .... The statements that the V cation is ``in
the almost perfect oxygen octahedron with some further distortions'' and
``The symmetry of the crystal field interaction \ldots\ is predominantly
octahedral'' and the caption to Fig. 1 are misleading. The oxygen
environment is severely distorted, which renders suspect any scheme based on
the disposition of atomic states in a perfect octahedral environment.'' Our
answer of 10.09.2002 '' for us your general objection ``why we use the
atomic picture'' as the start is curious as from two hundred years after
Davy, Dalton and others chemists all we know is that a solid is built from
atoms. Thus atomic starting point is simply obvious. It is not our problem
why other people in the field start in the very different way. Your
objection that we imply ``a strictly atomic scheme'' is not true. Also, that
we ``take no account of the milieu in which the V$^{3+}$ ions find
themselves in a V$_{2}$O$_{3}$ lattice'' -- is completely misunderstanding
of our job. We say just opposite -- the milieu, we use a more physical word
surroundings (both the charge and spin) is fundamentally important for the
realization in a solid of the given atomic-scale ground state and
consequently of the whole compound. It is a role of lattice distortions
widely discussed by us. In connection to distortions we fully agree that
``The oxygen environment is severely distorted'' but each consideration in
literature starts from the octahedron. Very important is that our electronic
structure, shown in Fig. 1c, will be only modified by distortions -- the
number of states will be preserved. Modification means the removal of
degeneracies shown, the change of the energy separations and shape of
eigen-functions. Next, our model provides both orbital and spin moment, Refs
8,13. We came out with our approach to 3$d$-atom containing compounds in
1996 when the orbital moment was not yet derived experimentally. Thus, we
take the possibility of accounting for the orbital moment as the great plus
for our theoretical approach.'' did not find the understanding of the
SCES-02 Committee. Surely, our many-electron CEF\ electronic structure is
completely different from the one-electron crystal-field electronic
structure with two electrons put on the t$_{2g}$ orbitals generally
discussed in the presently-in-fashion theories of 3$d$-atom containing
compounds.

The paper has been given under the law and scientific protection of the
Rector of the Jagellonian University in Krakow, of University of Mining and
Metallurgy and of Polish Academy of Sciences.

Figure caption:

Fig. 1. Electronic structure of the highly-correlated 3$d^{2}$ electronic
system occurring in the V$^{3+}$ ion calculated with the octahedral CEF
parameter $B_{4}^{0}$ = -40 K and the spin-orbit coupling $\lambda _{s-o}$%
=+150 K. According to the QUASST theory such the electronic structure is
expected to be largely preserved in a solid. In real V$_{2}$O$_{3}$ this
structure is slightly modified by off-octahedral distortions and magnetic
interactions.

\end{document}